\documentstyle[12pt,aaspp4,psfig]{article}
 
\received{2002 May 10}
\begin{document}

\title{HST Spectra of GW Librae: A Hot Pulsating White Dwarf in a Cataclysmic
Variable\footnote{Based on observations made with the NASA/ESA Hubble Space
Telescope, obtained at the Space Telescope Science Institute, which is
operated by the Association of Universities for Research in Astronomy, Inc.,
under NASA contract NAS 5-26555}}

\author{Paula Szkody} \affil{Department of Astronomy, University of
Washington, Seattle, WA 98195 
szkody@astro.washington.edu}

\author{Boris T. G\"ansicke} \affil{Dept. of Physics \& Astronomy, 
University of Southampton, Southampton SO17 1BJ, UK
btg@astro.soton.ac.uk}

\author{Steve B. Howell} \affil{Astrophysics Group, Planetary Science Institute,
Tucson, AZ 85705\\
howell@psi.edu}

\author{Edward M. Sion} \affil{Department of Astronomy, Villanova University,
Villanova, PA 19085
edward.sion@villanova.edu}

\begin{abstract}
We have obtained Hubble Space Telescope UV spectra of the white dwarf in GW Lib,
the only known non-radially pulsating white dwarf in a cataclysmic variable,
and the first known DAZQ variable. The UV light curve reveals
large amplitude (10\%) pulsations 
in the UV with the same periods (646, 376 and 237 s) as those seen at
optical wavelengths, but
the mean spectrum fits with an average white dwarf temperature
(14,700K for a 0.6M$_{\odot}$ white dwarf) that is too
hot to be in the normal instability strip for ZZ Ceti stars.
A better fit is achieved with a dual temperature model
(with 63\% of the white dwarf surface at a
temperature of 13300K and 37\% at 17100K),
 and a higher mass
(0.8M$_{\odot}$) white dwarf with 0.1 solar metal abundance.
Since the blue edge of the instability strip moves to higher temperature with
increasing mass, the lower temperature of this model is within the instability
strip.
However, the presence of accretion likely causes abundance and 
atmospheric temperature
differences in GW Lib compared to all known
single white dwarf pulsators, and the current models that have been capable
of explaining ZZ Ceti stars may not apply.
\end{abstract}


\section{Introduction}

With the discovery (van Zyl et al. 2000) of GW Lib as the 
first non-radially pulsating white dwarf
in a cataclysmic variable (CV), asteroseismology could
be applied, for the first time,
to understand the internal structure of an accreting white
dwarf. However, GW Lib does not easily relinquish its secrets. The six 
optical observing runs of van Zyl et al. (2002; VZ2002) 
showed that the pulsation spectrum was
highly unstable on timescales of months, which is common in cool,
hydrogen atmosphere white dwarf pulsators (ZZ Ceti stars or DAVs). 
While they could identify clusters
of signals with very close (2s) spacing, and some frequencies that 
were
repeatedly present in the optical runs (periods of 236, 376 and 648 s),
they could 
not disentangle the various
modes to interpret the pulsations. 
The presence of fine structure, linear combination modes and changes
in modes on monthly timescales all indicate that, in the optical at least, 
GW Lib is typical
of ZZ Ceti stars. However, cool pulsators usually have the largest
amplitudes. The relatively low amplitude of oscillation (5-17 mmag) in GW Lib
could be due to some dilution of the white dwarf optical light by an 
accretion disk.
VZ2002  concluded that they would need a
much longer baseline of data 
(e.g. Kleinman et al. 1998) to solve the problem.

Since GW Lib is too faint for a Whole Earth Telescope (Nather et al. 1990)
 campaign, a consortium of larger
telescopes will be needed to make progress from the ground. But 
the ultraviolet offers unique opportunities, as the white dwarf
usually contributes close to 100\% of the light in this portion of the spectrum
for low mass accretion rate systems (Szkody et al. 2002a; S2002a), and 
comparison
of the UV and optical amplitudes of pulsation can identify the modes of
DAVs (Robinson et al. 1995; R1995, Nitta et al. 2000; N2000). 
Thus,
GW Lib became
 an integral part of our Hubble Space Telescope study of white
dwarfs in short period dwarf novae.
This project uses the Space Telescope Imaging
Spectrograph (STIS) to obtain UV spectra which can be modelled to determine
the temperature, gravity, mass, rotation and composition (see G\"ansicke
et al. 2001; Howell et al. 2002, S2002a, Szkody et al. 2002b; S2002b for results). 

GW Lib is one of the WZ Sge type dwarf novae with 
very infrequent and
extreme amplitude outbursts (Howell, Szkody
\& Cannizzo 1995; HSC1995). Its only known outburst occurred in 
1983 (Duerbeck 1987).
Recent ground based time-resolved spectra
(Szkody, Desai \& Hoard 2000; SDH2000) 
revealed an orbital period near
79 min, one of the shortest of disk accreting CVs (Warner 1995). Subsequent 
data with longer time-coverage refines this period to 76.78 min
(Thorstensen et al. 2002). 
The optical spectra show broad absorption troughs (from the white dwarf)
 surrounding narrow
Balmer emission lines (from the low inclination disk), consistent with a low
mass transfer rate system. SDH2000 fit the absorption 
with an 11,000$\pm$1000K white
dwarf at a distance of 114 pc.
This temperature was within the general
location of the ZZ Ceti instability strip (11,200-12,900K; Koester \&
Holberg 2001, although the edges vary slightly with different authors) 
but dependent on the estimated disk contribution to the optical light. 
Surprisingly, the HST data reveal a much hotter white dwarf, as well as
the first known DAZQ variable. 

\section{Observations}

HST was scheduled to observe GW Lib for 4 successive orbits using
the STIS G140L grating and the 52 $\times$ 0.2 $\arcsec$ aperture to
obtain wavelength coverage from 1150-1720 \AA\ with a resolution of 1.2 \AA.
Problems with guide star or
target star acquisitions resulted in two failed
observations, but the third attempt on 2002 January 17 was
successful (Table 1).
The data comprise complete coverage of the binary orbit
 but are interrupted due to the low
earth orbit of HST. 
An optical spectrum obtained at Apache Point Observatory on January 9 showed
GW Lib was deep in quiescence, similar to the past data in SDH2000.

The data from the 4 orbits 
 were combined into a single spectrum (Figure 1).
This spectrum shows the typical broad
L$\alpha$ absorption from the high-gravity atmosphere of a white
dwarf, the quasi-molecular $H_2^+$ absorption at 1400\AA\ which is
indicative of temperatures below 18000K, as well as a whole range of
narrow photospheric low-ionization absorption lines (C\,I and II at
1280, 1335 and 1657\,\AA, OI at 1300\,\AA, Si\,II at 1260, 1300 and
1530\,\AA, and Al\,II at 1670\,\AA). One strong absorption feature
near 1355\,\AA\ remains unidentified. The presence of metals
means that the white dwarf in GW Lib is a DAZQ (Sion et al. 1983), and
different than ZZ Ceti stars which have pure H compositions.
It is noteworthy
that the $H_2$ absorption feature at 1600\AA\ that is typically present in
CV white dwarfs at temperatures below
13000K (S2002b) and which is evident in the the ZZ Ceti star
 BPM37093 (N2000, see Figure 1) is absent. 
In addition, the typical high-excitation emission lines observed in low
$\dot M$ CVs are present, even though relatively weak (He\,II at
1640\,\AA, C\,III and IV at 1175 and 1550\,\AA, N\,V at 1240\,\AA, and
Si\,III and IV at 1206 and 1400\,\AA).

The data were acquired in time-tag mode and a background-subtracted 
light curve was constructed using
all source photons except for a small region around L$\alpha$
(Figure 2). This light curve 
 shows evidence for strong and multiple periodicities. 

\section{Pulsation}

The Fourier transform of the light curve data
(Figure 3) shows significant power at 3 periods (646, 376 and
237 s with FWHM of $\sim$7 s). These periods are identical
 to those seen in the optical (VZ2002)
but the amplitudes in the UV are much higher (Table 2 summarizes 
the UV and optical pulses). 
According to Kleinman (1999), DAVs come in two period groupings; hot DAVs and
cool DAVs.
Both types show periods with $\sim$50 s spacing spanning 120 to 360
s (hot) and 300 to 650 s (cool). Only one DAV is known (G 29-38) which has
observed periods spanning the entire hot-cool range.
Kleinman presents a compilation of the available DAV periods together with 
the results for a 0.6 M$_{\odot}$ (hydrogen layer mass of 
10$^{-4}$ M$_{\mathrm{WD}}$) 
DAV model by Brassard et al. 
(1992) for a series of $\ell = 1$ modes.  The model periods generally match those
observed in both the hot and cool DAVs. 
The periods observed in GW Lib, like those in G 29-38, span the entire 
period range and approximately agree with both hot and cool DAV periods.
G 29-38 is a poorer match at the shortest (hot) periods while GW Lib does well
at both long (646 s) and short (236 s) periods and is within 2$\sigma$ of
the model period at 389 sec. 

R1995 and N2000 have previously used HST and
optical data together with
the fact that the pulsation amplitude ratio at different wavelengths is solely
a function of the spherical harmonic index $\ell$ to attempt mode determination 
for two single DAVs. From HSP
 filter observations at 1570 and 5500\AA, R1995 were 
able to use the observed pulse amplitude ratio of $\sim$6 for these wavelengths
to obtain a nice match to an $\ell = 1$ (dipole) mode model for the 
12500K, $\log g=8$ white dwarf in 
G117-B15A. 
N2000 used our instrumental setup of STIS with
G140L and found amplitudes near 1500\AA\ of 4-16 times the optical
amplitudes for the different pulsation periods in BPM 37093 (the most massive
 white dwarf pulsator known with $\log g=8.7$ and
T=11500K). While they could rule out $\ell\ge 3$ modes, the wavelength
dependence of the different periods did not allow a clear mode identification.
Although the best results are obtained for simultaneous
HST and optical observations (since the pulsation amplitudes are known
to vary in the optical; VZ2002), we can obtain some estimate for GW Lib from
the available non-simultaneous data, since the UV and optical periods are the 
same. 
Table 2 shows that the ratio of the UV/optical amplitudes ranges from
6-17, similar to the data on G117-B15A and BPM 37093,
 but contrary to the models
which show decreasing amplitude ratios for higher temperatures (R1995). 
Thus, at least in observed periods and the UV/optical amplitude
 ratio, 
GW Lib is similar to single pulsating white dwarfs.

\section{Spectral Fit}

Fitting white dwarf model spectra (Hubeny \& Lanz 1995) to the mean STIS 
spectrum
of GW\,Lib, assuming a canonical white dwarf mass of $M_{\mathrm
WD}\approx 0.6M_{\odot}$ ($\log g=8$) results in $T_{\mathrm
WD}=14700$\,K, 0.1 times solar metal abundances, and in a distance of
171 pc. Using 0.15 for the fraction of the accretion energy that would go
into heating the white dwarf (Sion 1985), the resulting accretion rate for this
temperature and white dwarf mass would be 4$\times 10^{-11}$ M$_{\odot}$/yr,
a value consistent with the low accretion rate of TOADs (HSC1995). 

While the fit is acceptable in the continuum and in the wings
of Ly$\alpha$, there are clear deviations around 1400\AA\ and in the
the core of Ly$\alpha$ (Figure 1). The quality of fit increases significantly 
($\chi^{2}$ decreases from 3.8 to 2.0) if a
two-temperature model is used, allowing for a variation of the
temperature over the white dwarf surface.
Again assuming $\log g=8$, the
best fit is achieved for $T_{\mathrm low}=13300$\,K and $T_{\mathrm
high}=17000$\,K, comprising 80\% and 20\% of the white dwarf surface,
respectively. While a dual temperature approach could
in principle explain the match of the periods of GW Lib to those
of both hot and cool DAVs, the most surprising result from these
spectral fits is that GW Lib is pulsing at all, as both $T_{\mathrm
low}$ and $T_{\mathrm high}$ are located out of the traditional
instability strip for ZZ Ceti stars!
It is perhaps
not so surprising that an accreting white dwarf should have an internal
structure different than single white dwarfs, but as noted earlier,
 the similarity of the periods
and UV/optical amplitude ratios does imply a similar pulsation mechanism
 and hydrogen layer 
mass in GW Lib and single DAVs.

Bradley \& Winget (1994) have explored the dependence of the blue edge of the
instability strip on stellar mass, on the hydrogen and helium layer masses and
on convection. While there was little change with H/He layer masses, 
they found
the blue edge of the instability strip moves to higher temperatures for
increasing convective efficiency and
increasing stellar mass. With their highest mass models (0.8M$_{\odot}$)
and efficient convection (ML3),
the blue edge reaches 13,500K. Thus, our 2 temperature white dwarf could
be in the instability strip if it has a higher mass than normal for single
DAVs. To explore this further, we fit our spectrum with an 0.8M$_{\odot}$
($\log g=8.4$)
white dwarf at a distance of 148 pc (Table 3). 
The 0.8M$_{\odot}$ model results in a hotter (15,500K) single
temperature white dwarf (which would remain outside the instability strip
unless the Bradley \& Winget model was extrapolated to the Chandrasekhar
limit), or a 2 temperature white dwarf with 63\% at 13,300K and 37\% at
17,100K (fit shown in Figure 4). 
Taking the results of the 0.8M$_{\odot}$ two temperature fit at face
value, the majority of the white dwarf surface has a temperature that
is within the ZZ Ceti instability strip.
Thus, a higher mass can resolve the dilemma of pulsation, but
the origin and effect of the dual temperatures remains to be explained. 
Rather than a multi-temperature surface, the dual temperatures may be related
to averaging over a temperature change during the pulsation.
Ongoing work on phase-resolving the time-tag spectra at the pulsation periods
should provide clues to the correct interpretation.

It is also possible that the dual temperatures
 could be due to a hot accretion spot on a magnetic,
spinning white dwarf. 
Using the Si 
and C lines at 1335, 1527, 1533 and 1657\AA, we find a lower limit on the
white dwarf rotation of $v\sin i<$
300 km/s, typical for the white dwarfs in CVs (S2002a). While this allows
us to rule out large rotation velocities, 
the low resolution of the G140L grating does not allow us to eliminate any of
the observed pulsation periods or fine structure (VZ2002) in the periods as
due to rotation.
Overall, the instability of the periods and the inability
to explain them all with spin/beat interactions does not support a
magnetic interpretation for the observed periods, but it is  
possible that there is some contaminating influence on the pulsation 
period structure
due to interaction of the binary and spin periodicities.
It will
certainly require further observations, especially at X-ray wavelengths, to
sort out this complication in addition to the usual pulsation structure.

\section{Conclusions}

Our STIS data on the only apparent non-radial pulsator in an accreting close
binary show that the pulsations in the UV are of large amplitude and have the 
same periods as seen in the optical. These periods match with those
evident in both hot and cool single non-radially pulsating DAs, implying
general similarities in structure to single ZZ Ceti stars, whereas the UV
spectrum indicates a DAZQ white dwarf with an inhomogenous temperature 
distribution over its
surface, or with a change in temperature during its pulsation.
The average 
UV/optical pulse amplitude ratio is similar to what has been
observed and predicted for DAVs in the $\ell = 1$ mode.
While the fit of the
spectrum to white dwarf models substantiates a dual-temperature structure
(13,300K from 63\% of the white dwarf surface and 17,100K from 37\% of the 
surface), 
both temperatures are hotter than evident for known single DAVs of 
0.6M$_{\odot}$. However, the cooler of the two temperatures 
is within the instability strip for more massive white dwarfs 
($\ge$0.8M$_{\odot}$). Thus, GW Lib may represent a system containing a more
massive and spotted white dwarf as compared to typical single DAVs. The mass
difference
may help to explain why other white dwarfs in CVs that contain similar 
or cooler temperatures than GW Lib are not pulsating -- e.g. HV Vir, EG Cnc,
VY Aqr (S2002b, Howell et al. 2002). 
The identification of GW Lib begs for time-dependent, non-adiabatic, non-radial
pulsation models of white dwarfs undergoing accretion at a low rate.

\acknowledgements
We gratefully acknowledge John Thorstensen for communicating his proper
motion measurement, which improved the HST acquisition of GW Lib, and
his improved orbital period.
Support for this work was provided by NASA
through grant GO-0813-97A from the Space Telescope Science Insitute,
which is operated by AURA, Inc., under NASA contract NAS5-26555. BTG was
supported by a PPARC Advanced Fellowship.

\clearpage

\begin{deluxetable}{lcc}
\tablenum{1}
\tablewidth{0pt}
\tablecaption{2002 January HST Data}
\tablehead{
\colhead{Orbit} & \colhead{UT Start} & \colhead{Time(s)} }
\startdata
1 & 01:52:00 & 2105.19 \nl
2 & 03:14:04 & 2603.19 \nl
3 & 04:50:15 & 2603.20 \nl
4 & 06:26 25 & 2580.20 \nl
\enddata
\end{deluxetable}

\clearpage
\begin{deluxetable}{lccc}
\tablenum{2}
\tablewidth{0pt}
\tablecaption{Pulse Periods and Amplitudes}
\tablehead{
\multicolumn{2}{c}{HST} & \multicolumn{2}{c}{Optical\tablenotemark{a}}\\
\colhead{P (s)} & \colhead{Amp (\%)\tablenotemark{b}} & \colhead{P (s)} & 
\colhead{Amp (\%)} }
\startdata
646 & 10 & 648 & 0.6-1.75 \nl
376 & 7 & 376 & 0.5-1.1 \nl
237 & 7 & 236 & 0.6-0.8 \nl
\enddata
\tablenotetext{a}{Values from VZ2002}
\tablenotetext{b}
{A=(2P$^{1/2})$/mean where A is amplitude and P is power spectrum}
\end{deluxetable}

\clearpage
\begin{deluxetable}{lccccc}
\tablenum{3}
\tablewidth{0pt}
\tablecaption{Model White Dwarf Model Fits to Data}
\tablehead{
\colhead{Model} & \colhead{log g} & \colhead{M$_{wd}($M$_{\odot}$)} &
\colhead{T (K)} & \colhead{\% Surface} & \colhead{$\chi^2$} }
\startdata
Single T & 8.0 & 0.6 & 14,700 & 100 & 3.8 \nl
2T       & 8.0 & 0.6 & 13,300+17,000 & 80+20 & 2.0 \nl
Single T & 8.4 & 0.8 & 15,500 & 100 & 3.6 \nl
2T       & 8.4 & 0.8 & 13,300+17,100 & 63+37 & 1.9 \nl
\enddata
\end{deluxetable}

\clearpage
%\begin{figure} [h]
%\figurenum {1}
\figcaption{Four orbit summed STIS spectrum of GW Lib with lines labelled. 
The 1$\sigma$
errors are shown at the bottom of the plot. The (arbitrarily
scaled) STIS spectrum
of the ZZ Ceti star BPM37093 from N2000 is plotted below GW Lib for comparison.
The light grey line
through the data is the best fit single temperature (14,700K) for a $\log g=8$
 white
dwarf.}

%\figurenum {2}
\figcaption{Light curve constructed from the time-tag data in 10 s bins for each of the 4 HST orbits with running
time from the start of the first orbit.}

%\figurenum {3}
\figcaption{Power Spectrum of the light curve data. The window function of the
3 identified periods is shown
at the top of the plot.}

%\figurenum {4}
\figcaption{Best two-temperature model fit to the data, for a $\log g=8.4$ 
(0.8M$_{\odot}$) white dwarf.
The two temperatures are 13,300K from 63\% of the white dwarf surface and 
17,100K from 37\% of the white dwarf. The emission lines
are approximated by Gaussians.}
%\end{figure}

%%%UCP%%%
\newpage
\plotone{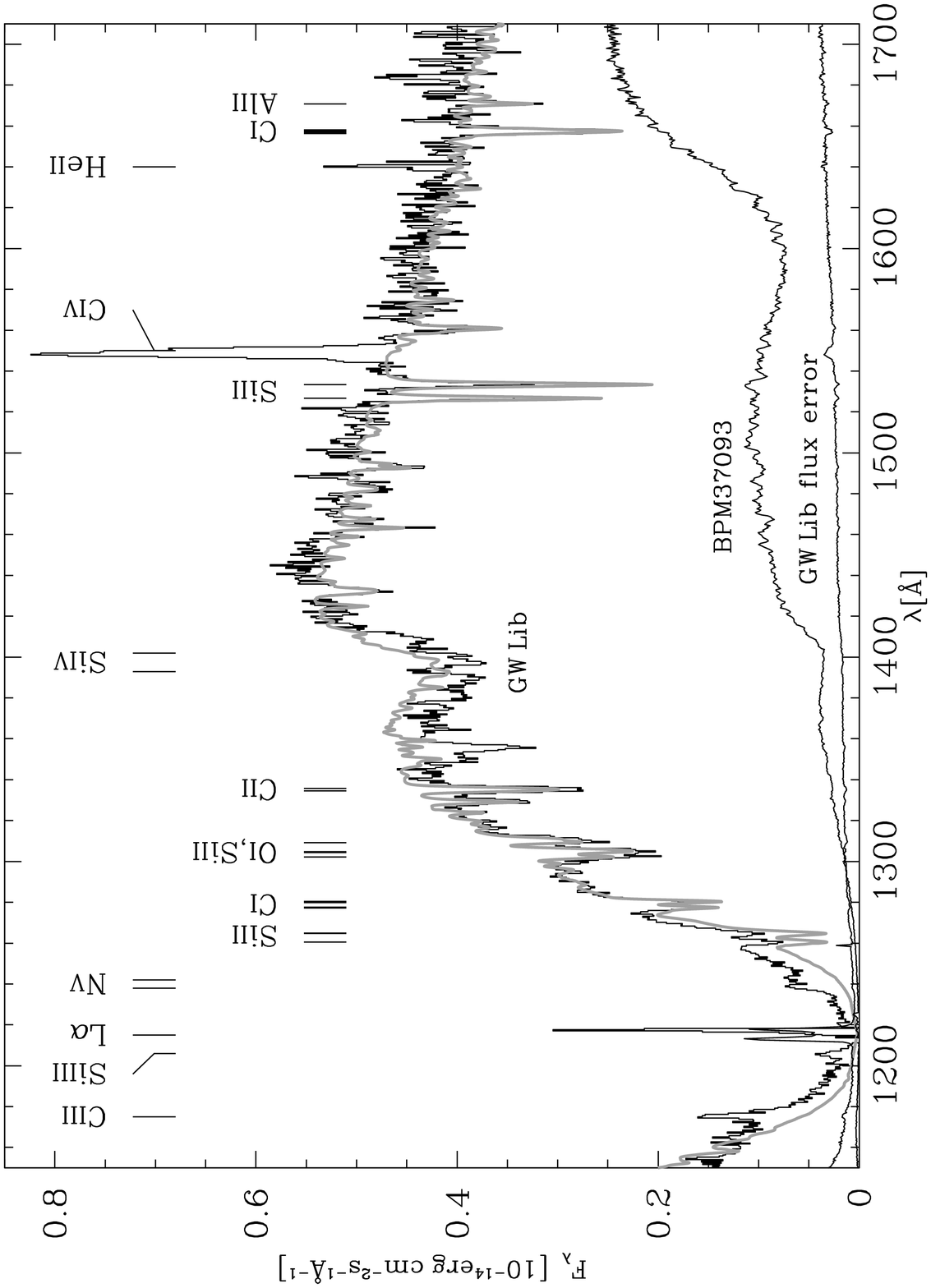}
\newpage
\plotone{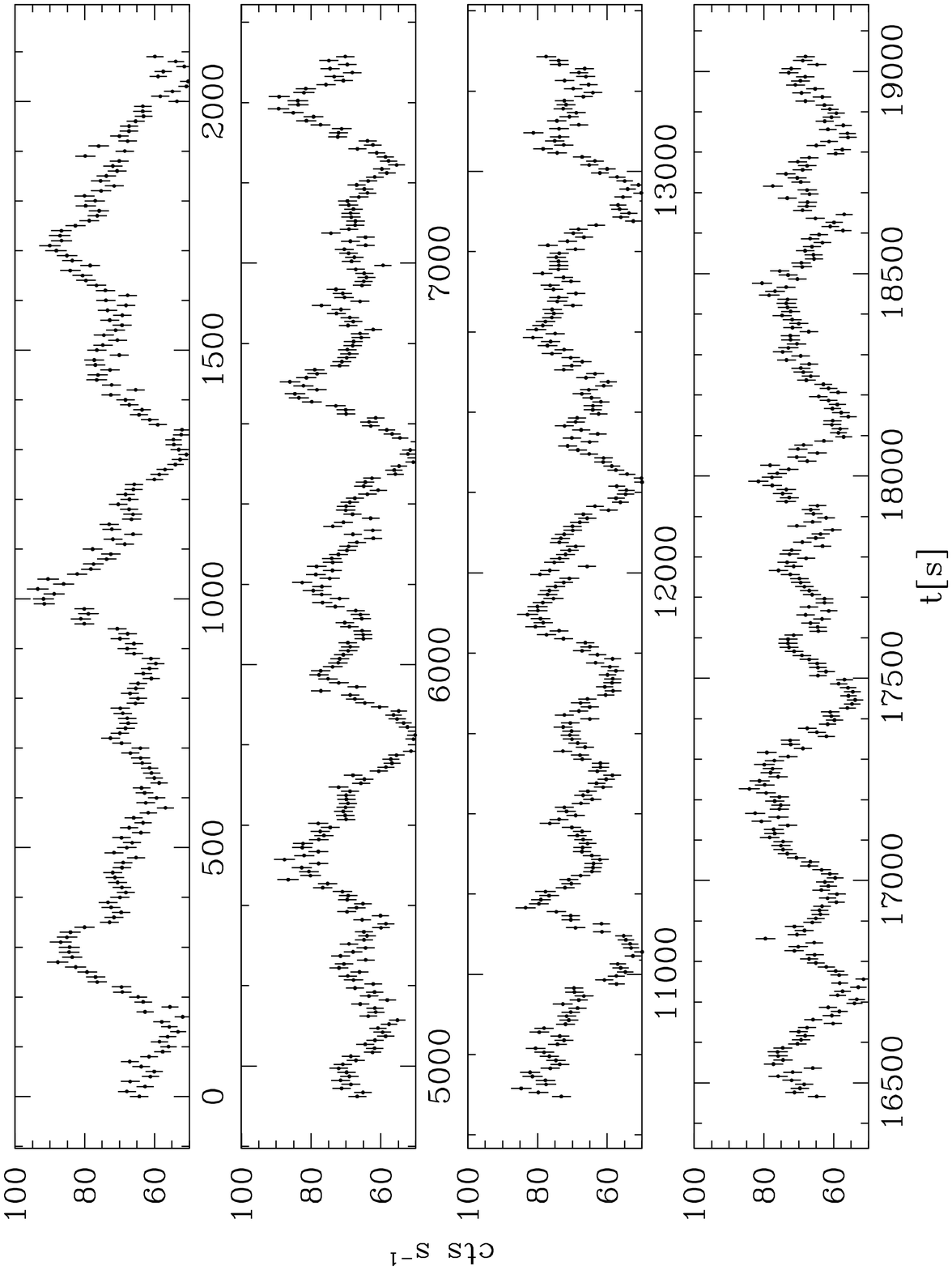}
\newpage
\plotone{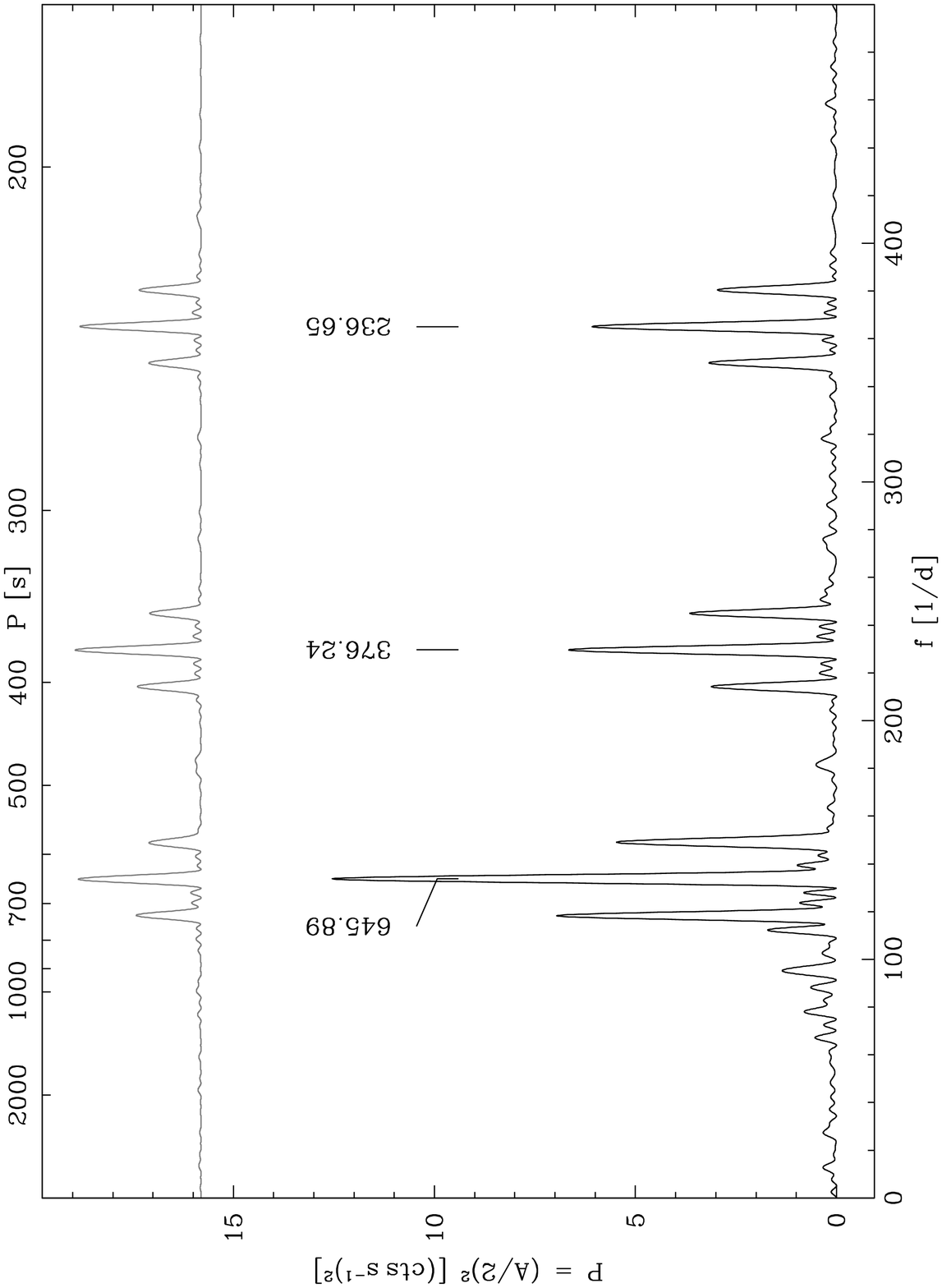}
\newpage
\plotone{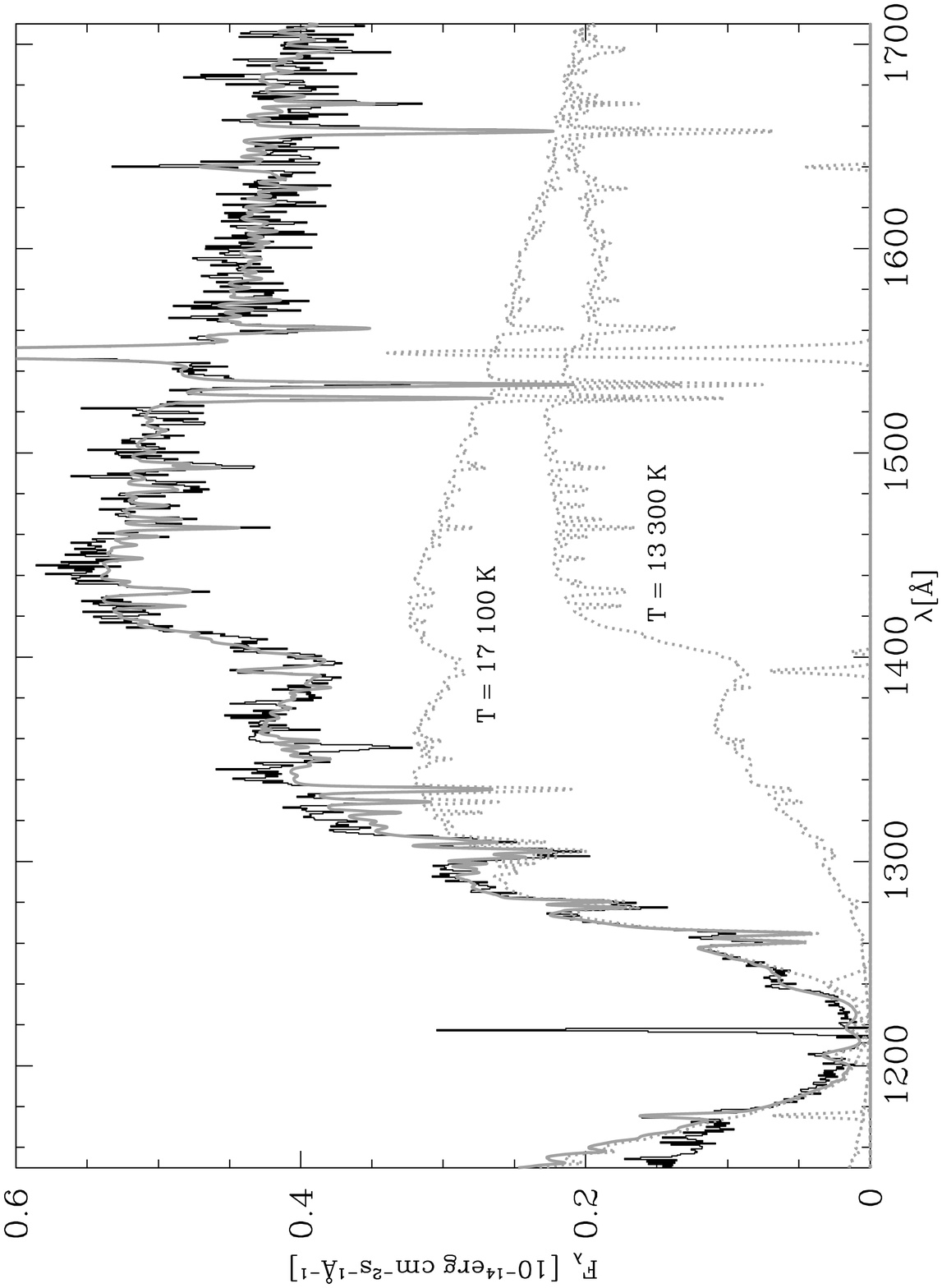}


\clearpage

\begin{references}
\reference{} Bradley, P. A. \& Winget, D. E. 1994, \apj, 421, 236
\reference{} Brassard, P., Fontaine, G., Wesemael, F, \& Tassoul, M. 1992,
\apjs, 81, 747
\reference{} Duerbeck, H. 1987, Space Sci. Rev., 45, 68
\reference{} G\"ansicke, B. T., Szkody, P., Sion, E. M., Hoard, D. W., Howell,
S. B., Cheng, F. H. \& Hubeny, I. 2001, \aap, 656, 661
\reference{} Howell, S. B., G\"ansicke, B. T., Szkody, P. \& Sion, E. M. 2002,
\apj, 575, in press
\reference{} Howell, S. B., Szkody, P. \& Cannizzo, J. 1995, \apj, 439, 337 (HSC
1995)
\reference{} Hubeny, I. \& Lanz, T. 1995, \apj, 439, 875
\reference{} Kleinman, S. J. 1999, in ASP Conf. Ser., 169, 11th European Workshop
on White Dwarfs, ed. J.-E. Solheim \& E. C. Meistas (San Francisco:ASP), 116
\reference{} Kleinman, S. J. et al. 1998, \apj, 495, 424
\reference{} Koester, D. \& Holberg, J. B. 2001, in ASP Conf. Ser., 226, 12th
European Workshop on White Dwarfs, ed. J. L. Provencal, H. L. Shipman, J.
MacDonald \& S. Goodchild (San Francisco:ASP), 299
\reference{} Nather, R. E., Winget, D. E., Clemens, J. C., Hansen, C. J., \&
Hine, B. P. 1990, \apj, 361, 309
\reference{} Nitta, A., Kanaan, A., Kepler, S. O., Koester, D., Montgomery, M. 
H. \& Winget, D. E. 2000, Baltic Astr., 9, 97 (N2000)
\reference{} Robinson, E. L., et al. 1995, \apj, 438, 908 (R1995)
\reference{} Sion, E. M. 1985, \apj, 297, 538
\reference{} Sion, E. M., Greenstein, J. L., Landstreet, J. D., Liebert, J.,
Shipman, H. L. \& Wegner, G. A. 1983, \apj, 269, 253
\reference{} Szkody, P., Desai, V. \& Hoard, D. W. 2000, \apj, 119, 365 (SDH2000)
\reference{} Szkody, P., G\"ansicke, B. T., Sion, E. M. \& Howell, S. B. 2002,
\apj, 574, in press (S2002b)
\reference{} Szkody, P., Sion, E. M., G\"ansicke, B. T. \& Howell, S. B. 2002,
in ASP Conf. Ser., 261, The Physics of Cataclysmic Variables and Related Objects, ed. B. T. G\"ansicke, K. Beuermann \& K. Reinsch (San Francisco:ASP), 21 (S2002a)
\reference{} Thorstensen, J. R., Patterson, J. O., Kemp, J. \& Vennes, S. 2002,
\pasp, in press
\reference{} van Zyl, L., Warner, B., O'Donoghue, D., Sullivan, D., Pritchard,
J. \& Kemp, J. 2000, Baltic Astr., 9, 231
\reference{} van Zyl, L., et al. 2002, \mnras, submitted (VZ2002) 
\reference{} Warner, B. 1995, in Cataclysmic Variable Stars (Cambridge, Cambridge
University Press)
\end{references}
\end{document}